# Optical rotation of ferrofluid on a horizontal substrate by Marangoni and thermomagnetic forces

Chengzhen Qin[1, 4, 5, †], Feng Lin[2, †, *], Laichen Liu[1], Chong Wang[2], Jiming Bao[3, 4, 5, *], Zhiming Wang[1, *]

[1]*Institute of Fundamental and Frontier Sciences, University of Electronic Science and Technology of China, Chengdu, Sichuan 610054, China*

[2] *School of Materials and Energy, Yunnan University, Kunming, Yunnan 650500, China*

[3]*Department of Electrical and Computer Engineering, University of Houston, Houston, Texas 77204, USA*

[4]*Materials Science and Engineering, University of Houston, Houston, Texas 77204, USA*

[5]*Department of Physics and Texas Center for Superconductivity, University of Houston, Houston, Texas 77204, USA*

†The authors contributed equally to this work.

*To whom correspondence should be addressed. Emails: fenglin@ynu.edu.cn, jbao@uh.edu, zhmwang@uestc.edu.cn

# Abstract

Light-actuation enables non-contact, precise, and flexible manipulation of liquids in microfluidics and liquid robots. However, it has long been a challenge to achieve optical rotation of liquid on a horizontal substrate because of the weak force of light. Here, we report, for the first time, the laser-induced rotation of macroscopic liquid placed on a horizontal substrate above cylindrical magnets. The investigated ferrofluid exhibits multiple spikes due to the Rosensweig phenomenon on the surface with different strengths of external magnetic field and rotates in a controllable direction by adjusting the laser beam position around ferrofluid spikes. This rotation results from the laser-induced displacement between the ferrofluid gravitational center, magnetic field center, Marangoni force, and non-uniform thermomagnetic force. Movement tracking of the spikes, thermal magnetization, and thermal imaging of ferrofluid were performed to elucidate the mechanism. Additionally, the influence of laser power, laser irradiation position, and magnetic field intensity on laser-induced rotation of ferrofluid are systematically studied. This work demonstrates a novel horizontal rotation of liquid using light and provides a deep understanding of the underlying mechanism and a flexible strategy for manipulating liquid for various optofluidic applications.

Keywords: Ferrofluid, Optofluidic, Magnetic field, Rotation, Fluid circulation

**Introduction:**

Manipulation of a droplet and driving liquid flow with light have been enormously studied and developed as a multidisciplinary field of microfluidics and optics, named optofluidics [1,2]. Because of the merits of non-contacting, high resolution, and flexible stimulation, optofluidic techniques were implemented in biochemical sensing [3,4], particles/cell manipulation [5-7], fluid control [8-11], and photo-energy conversion [12]. Based on the mechanism of radiation pressure [13], optical tweezer [14], opto-thermocapillary [5-7,11], photoacoustic streaming [9,15,16], asymmetric photothermal deformation [10] et al., liquid deformation, droplet moving and liquid flow have been realized with light. However, the light-actuation of liquid is limited to localized and miniature transport. It is still a challenge to achieve the macroscopic rotation and long-range directional circulation of fluid.

Magnetic field is another method that can realize macroscopic manipulation of magnetic liquid, such as ferrofluid or droplet with steel beads [17,18]. Ferrofluid, a suspension of magnetic nanoparticles in conventional solvents has been widely used in liquid seals, friction-reducing, medical devices, and fluid carriers since its invention in the 1960s [17,19-21]. The motion of the ferrofluid can be controlled by moving permanent magnets or tuning the electromagnetic matrix [17,18,21]. However, the operation precision and flexibility are far lower than the light actuation as it is difficult to build miniature magnets and control the motion of magnets accurately [17-19,21]. The combination of magnetic field and thermal demagnetization has long been used to drive the swinging and rotation of magnetic metals or solid magnetic composite, namely magnetic heat engine or Curie heat engine [22,23]. However, the manipulation of liquid by coupling laser and magnetic field, the corresponding properties and applications have never been studied before.

In this work, we demonstrate a series of new manipulations of ferrofluids with the combined drive of laser-induced thermal capillary and photothermal demagnetization of ferrofluids in magnetic fields. Under laser irradiation, the ferromagnetic droplet in the magnetic field rotates as a whole in the horizontal direction. By adjusting the laser irradiation position, the circular ferrofluid hill generated by the cylindrical magnet rotates clockwise and counterclockwise. And expanding the number of ferrofluid droplets from two to three and four can also be manipulated by lasers to rotate clockwise and counterclockwise. The combination of light and magnetic field to drive ferromagnetic fluid opens a new way for the basic and applied research of this technology in liquid drive, underwater communication, microfluidic switching and microfluidic power generation.

**Results and discussions**

In the absence of an applied magnetic field, ferromagnetic fluids are generally considered non-magnetic Newtonian fluids. However, when placed in a magnetic field, the ferrofluid is a Bingham plastic liquid due to the Rosensweig instability and forms protrusions[24].

To illustrate the combined action of laser and magnetic fields on ferrofluid, 650 microliters of ferrofluid were placed on a horizontally operated substrate. Adjusting the distance between the ferrofluid and the cylindrical magnet generated a double protrusion in the center of a glass substrate placed completely horizontally. The method of experimental operation is described precisely. Ferrofluid drops of known volume 650 μL are deposited on a horizontal substrate by using a micropipette and are subject to an external magnetic field. The diameter of a single Ferrofluid (FF) droplet is 5 mm. The light source is a CW laser with a wavelength of 532 nm. The magnetic field B is generated by a cylindrical neodymium magnet (diameter 25mm, thickness 5mm), of grade N35 and residual magnetic field 1.19 T. The diameter of the magnet is always much larger than the maximum size of the ferrofluid drops. This guarantees that the magnetic field can be considered homogeneous in the horizontal plane and only dependent on the vertical distance z between the magnet and the bottom of the drop.

As shown in Figure 1a, the movement of ferrofluid droplets is controlled by a laser, and the temperature change of the surface is captured by an infrared thermal imager. The magnetic fluid is a superparamagnetic fluid, which presents a magnetic mound under the distribution of magnetic induction lines in Fig. 1b, and the magnetic induction intensity distribution on the magnet surface presented by COMSOL simulation in Fig. c. Therefore, all magnetic flow droplets are stably distributed in the center of the magnetic field. The stable formation of the droplet is due to the magnetization against gravity. Ferrofluid can be affected by Marangoni force $F_M$[11,25], decay of magnetic susceptibility [26,27] and viscosity [28] properties under the combined action of magnetic field and laser. The optical image is shown in Fig. 1 (d) -(f). ferrofluid robots are manipulated by laser fingers to perform a series of magical movements.

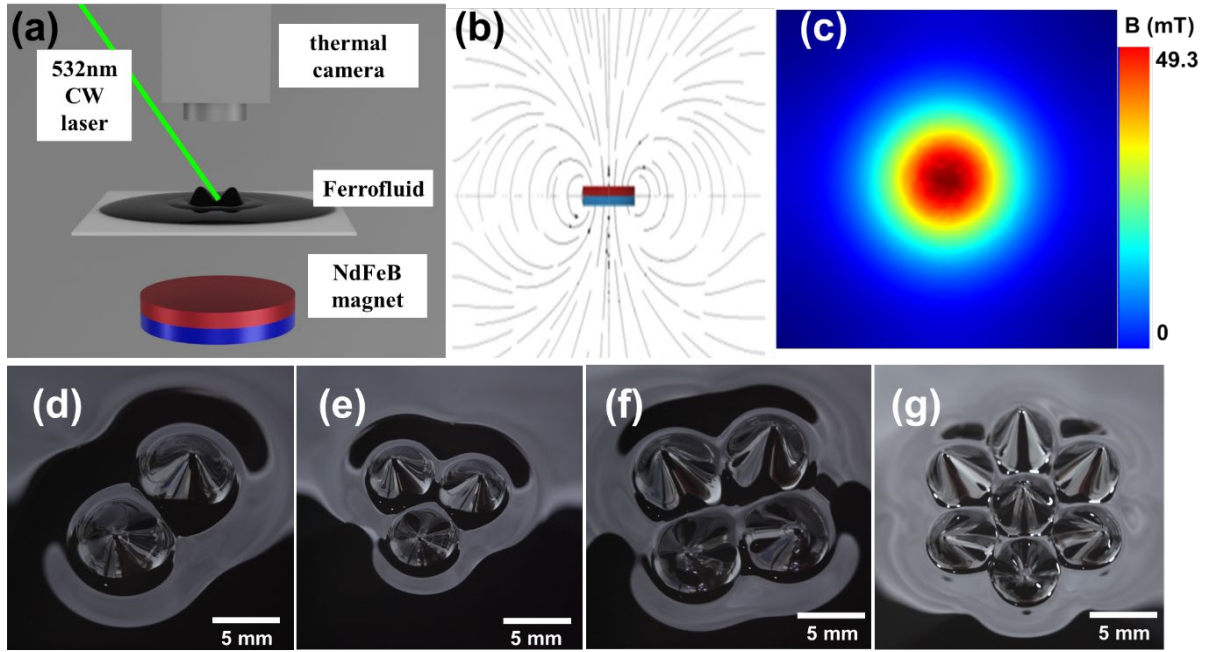

Figure 1. Under the action of the magnetic field, droplet ferrofluid can be controlled into any shape (a) (a) Schematic diagram of the experimental device, in which the laser finger manipulates the motion of the magnetic fluid droplets with a power of 400mW. (b) Magnetic induction line distribution of the magnet (c) magnetic induction intensity distribution (d)-(g) 650 microliter ferrofluid under the magnetic field of a cylindrical NdFeB permanent magnet produces different forms of protrusions in the magnetic field center under the Rosenweig effect.

The manipulation and understanding of individual droplets open the possibility of using lasers to manipulate larger numbers of FF droplets. By adjusting the distance between the NdFeB cylindrical permanent magnet and the ferrofluid, a single ferrofluid droplet is transformed into two ferrofluid mounds.

The motion of the ferrofluid mound under the effect of magnet rings and the laser is shown in Figs. 2a-b. Taking two bulges as an example, one of the bulges was selected and marked as a reference droplet with a red mark. By irradiating a 300mW laser on a symmetric position, it could be manipulated to exhibit two states of clockwise and counterclockwise motion.

When the laser is irradiated in the middle of the dome but deviates from the axis of symmetry, the ferrofluid protrusions can show different directions of motion according to the position where the laser deviates from the axis of symmetry, as shown in Figs. 2a-b. In schematic Figure 2a, the photothermal effect in the middle of the laser leads to attenuation of the magnetic attraction $F_m$, and the Marangoni force $F_M$ on the side irradiated by the laser is stronger, producing a corresponding clockwise flow. Surprisingly, a clockwise or counterclockwise rotating flow was observed in Figs. 2a-b when the dome of ferrofluid was irradiated either on the upper or lower side (SV1 SV2). This is the first demonstration of controlling liquid rotation in the horizontal direction. Rotation adjustment is discussed below. It is worth mentioning that the rotation of the ferrofluid is driven by the laser irradiation, not the movement of the magnet because the rotation of the magnet does not affect the rotation of the ferrofluid.

As shown in Fig. 2a, a longitudinal axisymmetric line is drawn for two ferrofluid droplets. At 0 s, the laser is added, and the position is above the ferrofluid demarcated by the red inverted triangle. At this time, a red dotted line can be artificially set to demarcate the position of the ferrofluid. The ferrofluid droplet labeled by the triangle moved 1 mm to the right. At this time, the ferrofluid was no longer in the center of the magnetic field, and the force it received was unbalanced. After 3 s, the ferrofluid began to rotate clockwise, and 10 s was the schematic diagram of two ferrofluid droplets turning half circle. 18 s shows a snapshot of a ferrofluid droplet rotating clockwise for one turn. In contrast, for Fig. b, the anti-clockwise rotation of the two ferrofluid droplets can be observed simply by moving the laser from above to below the axis of symmetry. Similarly, at 1 s, a shift occurs in the ferrofluid. After 8 s, the ferrofluid begins to rotate. 15 s is the schematic diagram of the ferrofluid rotating half a circle counterclockwise, and 28 s is the ferrofluid rotating back to the origin.

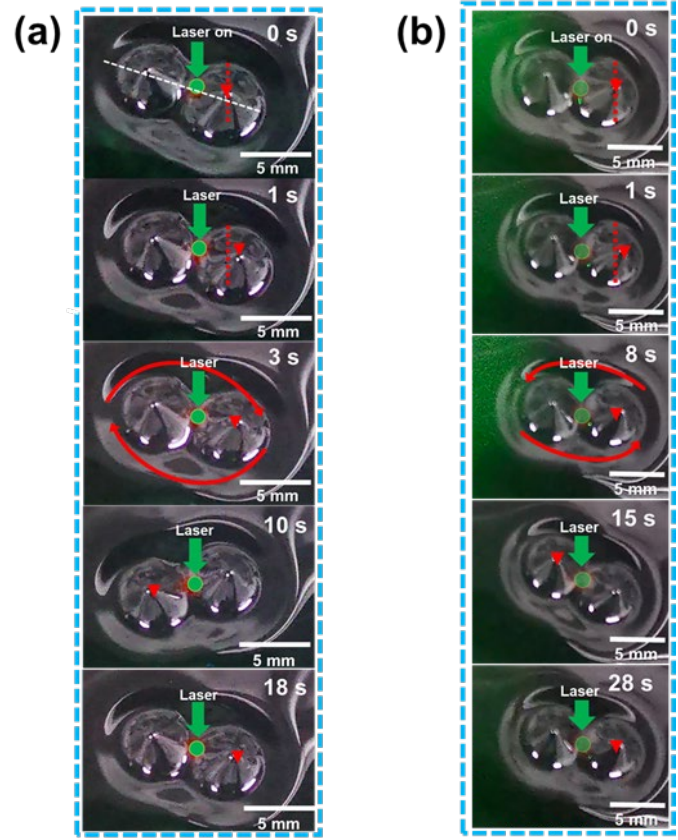

Figure 2. The laser spins two ferrofluid droplets clockwise and counterclockwise. (a) Laser manipulation of two magnetic liquid droplets clockwise for a week. Laser is added at 0s, and the magnetic fluid shifts in the magnetic field due to thermal demagnetization after 1s. After 3s, the FF starts to rotate clockwise, turns half a circle in 10s, and completes a circle in 18s. (b) The laser manipulates the two ferrofluid droplets to complete a counterclockwise rotation. Laser is added at 0s, and the magnetic fluid shifts in the magnetic field due to thermal demagnetization after 1s. After 8s, the FF starts to rotate counterclockwise, turns half a circle in 15s, and completes a circle in 28s

Figs. 3a-c shows the experiment schematic with a horizontal incident laser with two bulges as an example, one of the bulges was selected and marked as a reference droplet with a red mark. By irradiating a 300mW laser on a symmetric position, it could be manipulated to exhibit two states of clockwise and counterclockwise motion. The formula of Kelvin force is

$$F = \mu_0 MH$$

$$M = M_0 - k\Delta T$$

The direction of F is the same as the direction of the H gradient. When the right side is heated, the temperature increases, ΔT increases, and F on the right side decreases, and the overall force of FF moves to the right. In Fig. 3a, no laser is added, the center of gravity of the ferrofluid(COGFF) and the center of the magnetic field(COMF) coincide, and the force is balanced. In Fig. 3b, When the laser irradiates the upper right of the FF symmetry axis, the FF is unevenly heated, and the FF shifts to the right. At this time, the direction of the body force (volume force) $F_K$ is along the diagonal direction and points to the "first quadrant", shown in Fig 3c. Because the COGFF and the COMF do not coincide, the FF as a whole receives a component force to the right and finally realizes clockwise rotation. In contrast, when we use lasers to manipulate FF droplets to rotate counterclockwise. When the laser irradiation position is in the "fourth quadrant", the direction of the body force (volume force) $F_K$ is along the diagonal direction and points to the "fourth quadrant", and the horizontal tangential component forces drive the FF to rotate counterclockwise. Based on the theoretical analysis of thermal demagnetization, the magnetization of the required ferrofluid was tested with the change of temperature in the experiment. The test results are shown in Fig. 3d. It can be seen that with the increase in temperature, the magnetization tends to decline. Therefore, in the process of laser irradiation, the temperature rise caused by the photothermal effect plays a crucial role in the control of ferrofluid motion.

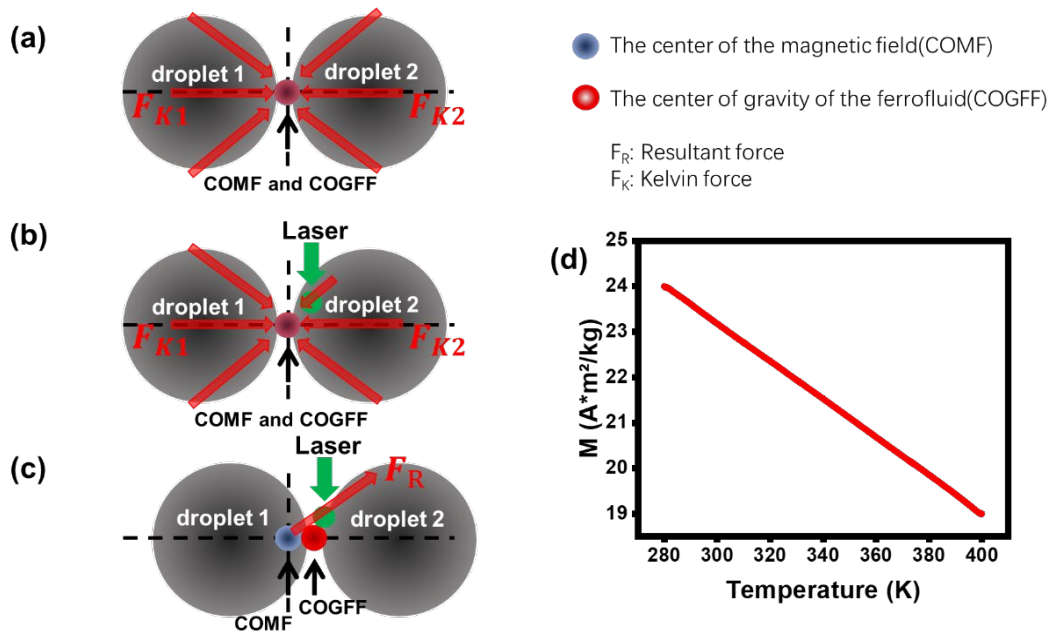

Figure 3. Schematic diagram of horizontal migration principle of laser-driven

Ferrofluid. (a) Magnetohydrodynamic force balance. (b) The photothermal effect causes the imbalance of magnetic fluid force. (c) The magnetic fluid deflects under the action of force. (d) Scatter plot of magnetization of magnetic fluid EFH1 as a function of temperature.

Moreover, we attempted to drive the ferrofluid droplets by direct irradiation with the laser light of 532 nm to the center of the ferrofluid droplets. In the experiment, the rotation direction of the ferrofluid can be controlled by adjusting the irradiation position of the laser. It is worth mentioning that once the laser starts to irradiate the magnetic fluid, the position of the laser, the size of the magnetic field, and other experimental parameters will always remain unchanged, we only need to wait for the magnetic fluid droplets to keep continuous rotating motion under the continuous irradiation of the laser. Based on the basic research on the rotational motion of FF in the vertical direction[29], we can apply this theory to explore the strange physical phenomenon of horizontal rotation. As shown in Fig. 4, in the case of moving the light source in a symmetric direction, the ferrofluid droplets as a whole can be driven in opposite directions clockwise and counterclockwise (movies S1 and S2). Such a noticeable phenomenon can be explained as follows. (i) When the laser is added, the magnetic susceptibility of the ferrofluid around the irradiated site increases with increasing temperature due to the photothermal effect. Because the laser heating position is biased to the right, the Kelvin Force of magnetic fluid is unbalanced, and $F_{k2}$ becomes smaller and $F_{k1}>F_{k2}$, so FF is shifted to the right as we described before in Fig 3. (ii) Since the position of the laser is unchanged, the position of the ferrofluid bulge relative to the laser is different from the initial position. (iii) Due to the strong magnetic field in the center, the magnetic fluid that has deviated from the center of the magnetic field will be "pulled" by the center of the magnetic field, and (iv) then the low temperature above it will be stronger than the lower magnetic field force, and the pulling force of the center of the magnetic field will be different, so the force imbalance is shown as $F_2>F_1$; (v) the overall trend shows counterclockwise rotation in Fig. 3c.

By contrast, by moving the laser from the "first quadrant" to the "fourth quadrant", the ferrofluid's force is reversed, and the whole body appears to move counterclockwise. Repeating these processes enables motion control of the FF droplets. Figure 4 illustrates a schematic diagram of the principle of controlling the rotational motion of a ferrofluid in the opposite direction.

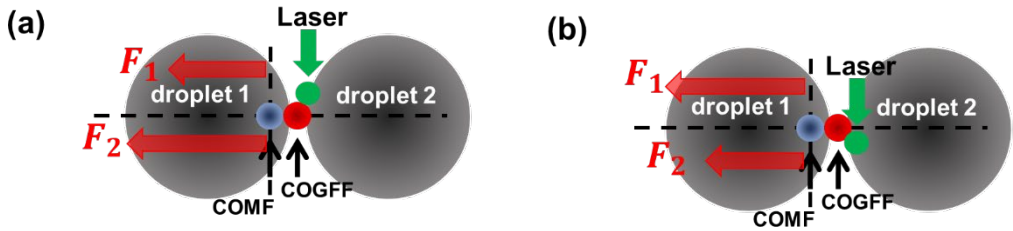

Figure 4. Rotation analysis diagram of the rotating motion of two ferrofluid droplets. (a) Horizontal clockwise rotation of the force diagram. (b) Horizontal clockwise rotation of the force diagram.

The photothermal response of the ferrofluid was evaluated by infrared thermography. The 532 nm CW laser was aimed at the center of two droplets. Thermograms were recorded using the function of an infrared thermal imager to shoot video (SV3 SV4). A thermal image video of the entire process was recorded from the addition of the laser to the completion of the circular rotation of the two ferrofluid droplets.

To see the heat distribution of the surface and to better observe the microscopic situation, We took thermal imaging pictures of the magnetic fluid under 300 mW laser irradiation. The thermal image of two ferrofluid droplets rotating in a counterclockwise rotation is shown in Figure 5, and the thermal image of a clockwise rotation corresponds to Figure 6.

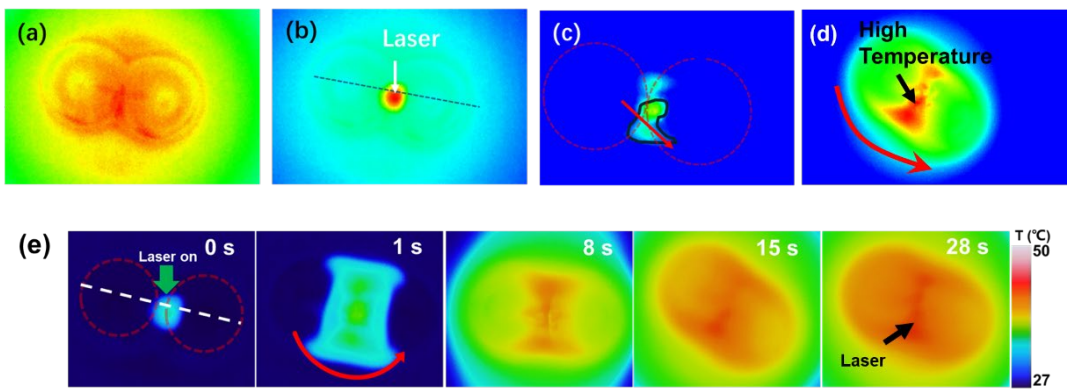

Figure 5. Thermal image of the Counterclockwise rotation of two ferrofluid droplets. (a) thermal image without the laser, (b) the laser shines on the lower right of the FF, (c) the heat distribution at the beginning, and (d) (e) is the heat distribution during rotation.

Figure 5 (a) shows that the FF is static in the horizontal plane at the beginning, and then the laser shines on the lower right of the FF. As shown in Figure 5(b), the temperature difference will cause the FF to have an uneven force, thus disturbing the equilibrium state of the FF. As shown in Figure 5 (c), the heat of the FF distributes faster along the red arrow, and the FF shifts to the right. As the laser is irradiated on the left side of the magnetic field center and the FF's center of gravity, the FF rotates counterclockwise. Fig.5 (d) shows a frame during FF rotation. The temperature in the red region of FF is the highest, and the laser continues to act on it, so FF rotates counterclockwise. Fig.5 (e) shows the surface temperature distribution diagram of FF rotating nearly one circle. After 28s, the ferromagnetic flux will return to the original position. Due to the heating action of the laser, the ferrofluid continues to rotate counterclockwise.

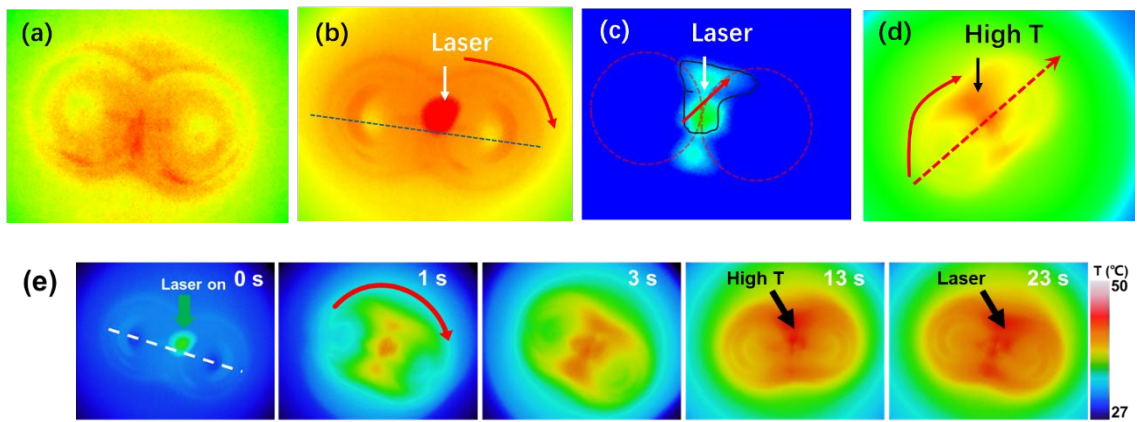

Figure 6. Thermal image of the clockwise rotation of two ferrofluid droplets. (a) thermal image without the laser, (b) the laser shines on the higher right of the FF, (c) the heat distribution at the beginning, and (d) (e)the heat distribution during rotation at the end.

Similarly, when the laser is applied to the upper right side of the FF droplet symmetry, as shown in Fig.6 (b), the laser can drive the FF to produce a clockwise rotational motion. As shown in Figure (b), the position of the laser is on the right side of the line between the FF center of gravity and the magnetic field center. According to the theory mentioned earlier in the article, the uneven force caused by the uneven heat distribution caused by the irradiation on the top will make FF produce the effect of clockwise rotation. The initial heat expands most rapidly along the direction of the red arrow in Fig.6 (c), making it produce the force effect shown in Figure 4a. Fig.6 (d) shows a frame in which the FF almost returns to its original position after one clockwise rotation. As shown in the figure, the high-temperature heat is distributed to the right of the line between the magnetic field center and the FF center of gravity, and the laser drives the FF to continue to rotate clockwise until the laser is withdrawn.

Finally, we have experimentally demonstrated the possibility of developing more droplet systems. Intriguingly, the FF above 'cylinder' type magnets (Fig. 7) begin to rotate during light irradiation at the edge of the droplet (SV5 SV6 SV7 SV8). The rotational speed was estimated to be around 3 rpm when the laser power was 500 mW. Judging from the infrared thermography images for the rotating ferrofluid system, it is considered that the distorted distribution of the temperature of the ferrofluid triggers rotatory motion. By applying the theoretical and practical basis of driving two ferrofluid rotating motions, as shown in Figure 7, we can easily achieve the practical application of manipulating multiple ferrofluid motions. Fig. 7c presents a method to drive three ferrofluid droplets to rotate horizontally and counterclockwise. When a continuous laser at 532 nm irradiates the left side of the red mark, the ferromagnetic fluid absorbs energy and generates a local temperature difference, resulting in the thermal demagnetization effect and the Marangoni effect. The droplet of ferrofluid begins to rotate counterclockwise as a whole. The location of the laser light is more favorable to the reference ferrofluid we selected than to the other two. Secondly, the position of the left side of the droplet irradiated by the laser (relative to the axis of symmetry), it is not difficult to analyze the principle of the three motion modes by referring to the two ferrofluid motion models. Figure 5a shows the actual picture driving the horizontal counterclockwise rotation of the three ferrofluid droplets. 0 s is the schematic diagram of laser addition, and the next 5 s, 10 s, and 15 s are the actual pictures of the ferrofluid counterclockwise process (refer to the position marked by the red triangle). After 15 s, the droplets return to the initial position. The laser manipulates three ferrofluid droplets to complete a counterclockwise rotation. Compared with the counterclockwise movement, the ferrofluid absorbs energy to produce a local temperature difference, and the droplet rotates clockwise as a whole. 17 s after rotation. Figs. 5c-d shows the actual picture of the horizontal counterclockwise and clockwise rotation of the four ferrofluid droplets driven. 0 s is the schematic diagram of the addition of the laser. After 1 s, the ferrofluid has an off-center shift, and the position of the ferrofluid in the magnetic field is no longer in the center, and the magnetic field force it receives is also uneven. After 2 s, the four droplets began to rotate counterclockwise as a whole. The red triangle marks in the figure are used to label a ferrofluid droplet for convenient description and observation of its movement. After 40 s, the droplet turned to a diagonal position, and after 90 s, the droplet returned to the initial position. The laser controlled the four ferrofluid droplets to complete a counterclockwise rotation. The picture in Fig. 5d is a real shot of the horizontal clockwise movement of the four ferrofluid droplets driven by the laser. The figure shows 90 s snapshots of the time from the time the laser is added to the start of the rotation and the last turn of forwarding.

When the magnetic field intensity was weaker than 18.3 mT, there was no ferrofluid bulge. With the increase of the magnetic field intensity, a ferrofluid mound was generated at 19.4 mT. Then a one-dimensional platform was used to precisely adjust the distance between the ferrofluid and the horizontal operating platform. Fig. 7e shows the relationship between the number of ferrofluid droplets and the magnetic field intensity. The relationship between laser power and fluid velocity and the relationship between the number of ferrofluid droplets and fluid velocity under the same power was tested by adjusting the power. Fig. 7d data show that the liquid rotation speed increases with the increase of power, whether it is three or four ferrofluids, in rad/min. It is worth mentioning that the smaller the quantity of ferrofluid liquid, the more its speed increases with the increase of power. Conversely, the speed increase rate of the four ferrofluid liquid with larger mass is smaller than that of the three ferrofluid liquid droplets with smaller mass with the increase of power. Therefore, the more ferrofluid droplets there are at the same power, the slower the speed will be. It is well known that the magnetic field will increase the viscosity of Bingham plastic ferrofluid [24,30,31]. The low moving resistance coupling with attenuated magnetic attractive force leads to the enhancement of rotation speed. The reduced magnetic force and enhanced Marangoni convection contribute to the rapidly increasing rotation speed in Fig. 7f.

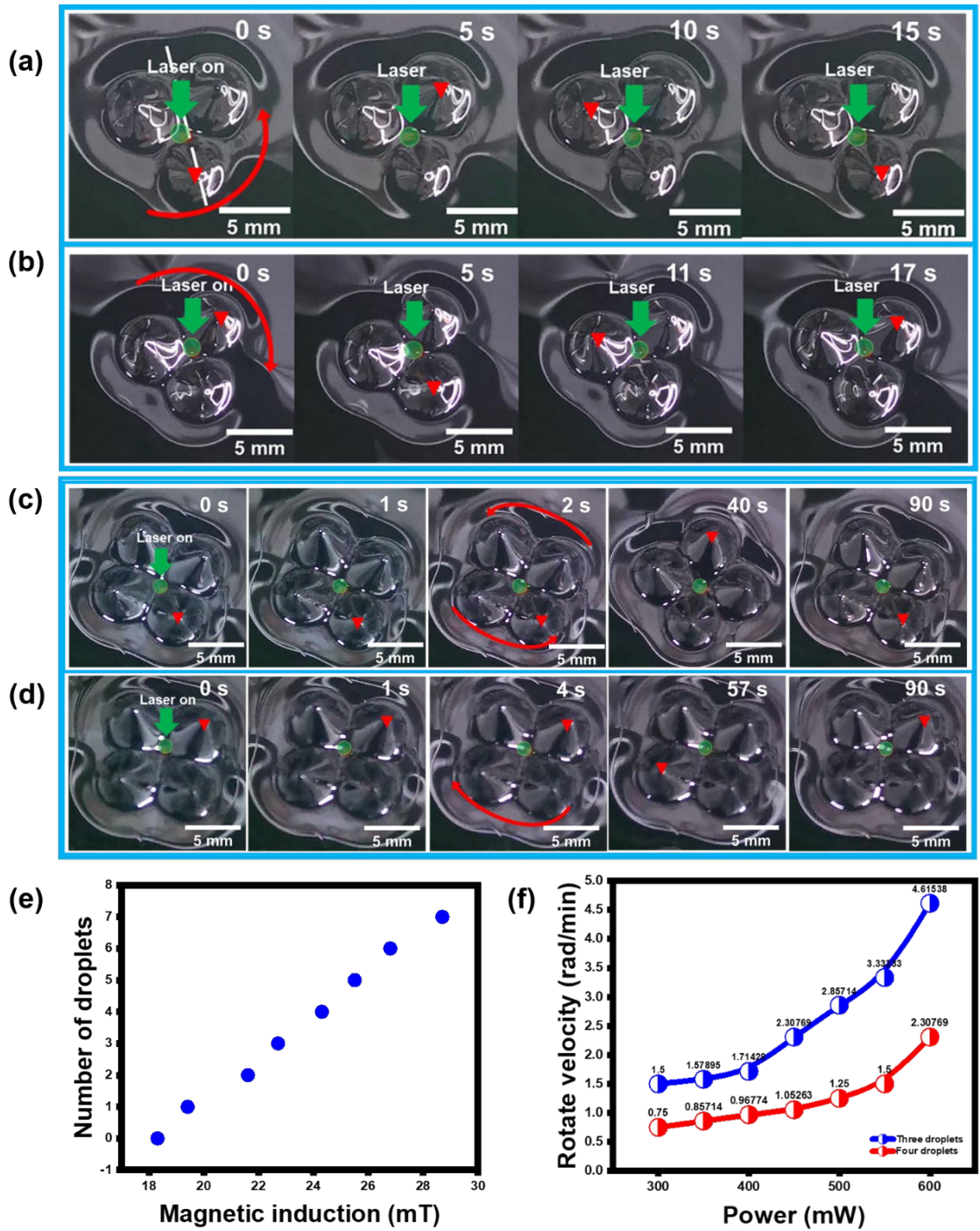

Figure 7. A 500 mW laser can also manipulate multiple ferrofluid drums to achieve rotation in different directions. (a) Laser manipulation of three ferrofluid droplets to complete a counterclockwise rotation for 15 s.(b) Laser manipulation of three ferrofluid droplets for 17 s to complete a clockwise rotation. (c) Laser manipulates four ferrofluid droplets for 90 s to complete a counterclockwise rotation. (d) Laser manipulates four ferrofluid droplets 90S to complete a clockwise rotation. (e) Graph of the number of the magnetic droplet and magnetic field strength. (f) Diagram of rotation speed and power.

Based on the macroscopic rotation operation of ferrofluid, it has great guiding significance for macroscopic application. For conventional driving of ferrofluid, it is always necessary to move the magnets or control the electromagnets matrix [17-19,21]. Here, the macro control we demonstrate does not require any additional mechanically driven pumps. The thermomagnetic convection of ferrofluid has been expected in heat dissipation of electronic devices for the increasing integration density and power consumption [20]. Compared with the coolants of oil or water, nanofluid exhibits better performance because of the high thermal conductivity of nanoparticles inside [32].

In addition, the use of ferrofluid as a carrier for carrying solid objects offers the possibility for the development of multifunctional robots. In addition to rotating motion, it was found that the laser can control the rocking motion of the ferrofluid, which will provide future applications such as radiators, underwater ultrasonic communications, photothermal mechanical transducers, and optical fluid devices [18,21,23,33].

**Discussion**

In summary, under the action of the laser, the mechanical equilibrium of the ferrofluid is broken by the photothermal effect, the heat convection of the fluid is induced by the laser, and the macroscopic spin motion of the fluid is further realized, which proves for the first time the spin motion of the ferrofluid droplets in the horizontal direction without the action of macroscopic gravity. The ferromagnetic fluid selected in this work is a material with strong light absorption and the surface tension coefficient varies greatly with temperature. Cylindrical magnets are used to stimulate surface instability and produce different amounts of droplets. We further found that higher laser power and a smaller number of droplets are both beneficial for improving the rotation speed. At the same time, such spins need to be verified in a rotationally symmetric magnetic field. It is worth mentioning that this is a large-scale horizontal macro-control. The combination of magnetic field and photothermal effect not only provided an effective tool for complex and flexible manipulation of ferrofluid but also a liquid operation platform. This work opened up the fundamental research on the actuation of liquid with multiple effects of the magnetic field, and broad applications such as optofluidics, microfluidic switch, microfluidic communication and heat dissipation.

**Methods**

Material

The oil-based ferrofluid (EFH1) were purchased from Ferrotec Corporation. The physical parameters of the material have been evinced in the supplementary information.

Experiment and measurement

The thermal images and temperature of the ferrofluid were obtained using a thermal camera (FLIR A655sc). The magnetic field was measured with a Gaussmeter (WT10A, WEITE MAGNETIC, China). The magnetization curve of oil-based ferrofluid was measured by a Vibrating Sample Magnetometer (Lake Shore Cryotronics, 8604-SSVT). The magnetic field was generated by horizontally placed cylindrical NdFeB magnets. The strength of the magnetic field was controlled by adjusting the space between the magnets and ferrofluid. The number of ferrofluid droplets was controlled by the strength of the magnetic field. The temperature of the ferrofluid was also measured using a thermal camera (FLIR, A655sc). By marking the rotation position of a single ferrofluid droplet, the rotation velocity of the ferrofluid under different laser power, different numbers of ferrofluid droplets, and different laser irradiation positions were measured.

**Data and materials availability**

The data that support the findings of this study are available in the supplementary material of this article.

**Declaration of competing interest**

The authors declare that they have no known competing financial interests or personal relationships that could have appeared to influence the work reported in this article.

Acknowledgments

F. L. acknowledges support from National Science Foundation of China (No. 52002049) and the International Postdoctoral Exchange Fellowship Program (No. 2020061). Z. M. W. acknowledges the National Key Research and Development Program of China (2019YFB2203400), National Science Foundation of China (No. 62075034), and the "111 Project" (B20030).

**REFERENCES**

Uncategorized References

1       Psaltis, D., Quake, S. R. & Yang, C. Developing optofluidic technology through the fusion of microfluidics and optics. *Nature* **442**, 381-386, doi:10.1038/nature05060 (2006).
2       Monat, C., Domachuk, P. & Eggleton, B. Integrated optofluidics: A new river of light. *Nat. Photonics* **1**, 106-114, doi:10.1038/nphoton.2006.96 (2007).
3       Yanik, A. A. *et al.* An optofluidic nanoplasmonic biosensor for direct detection of live viruses from biological media. *Nano Lett.* **10**, 4962-4969, doi:10.1021/nl103025u (2010).
4       Fan, X. & White, I. M. Optofluidic Microsystems for Chemical and Biological Analysis. *Nat. Photonics* **5**, 591-597, doi:10.1038/nphoton.2011.206 (2011).
5       Hu, W., Fan, Q. & Ohta, A. T. An opto-thermocapillary cell micromanipulator. *Lab Chip* **13**, 2285-2291, doi:10.1039/c3lc50389e (2013).
6       Mallea, R. T., Bolopion, A., Beugnot, J.-C., Lambert, P. & Gauthier, M. Laser-Induced Thermocapillary Convective Flows: A New Approach for Noncontact Actuation at Microscale at the Fluid/Gas Interface. *IEEE ASME Trans. Mechatron.* **22**, 693-704, doi:10.1109/tmech.2016.2639821 (2017).
7       Winterer, F., Maier, C. M., Pernpeintner, C. & Lohmuller, T. Optofluidic transport and manipulation of plasmonic nanoparticles by thermocapillary convection. *Soft Matter* **14**, 628-634, doi:10.1039/c7sm01863k (2018).
8       Liu, G. L., Kim, J., Lu, Y. & Lee, L. P. Optofluidic control using photothermal nanoparticles. *Nat. Mater.* **5**, 27-32, doi:10.1038/nmat1528 (2006).
9       Wang, Y. *et al.* Laser streaming: Turning a laser beam into a flow of liquid. *Sci. Adv.* **3**, e1700555, doi:10.1126/sciadv.1700555 (2017).
10      Lv, J. A. *et al.* Photocontrol of fluid slugs in liquid crystal polymer microactuators. *Nature* **537**, 179-184, doi:10.1038/nature19344 (2016).
11      Gao, C. *et al.* Droplets Manipulated on Photothermal Organogel Surfaces. *Adv. Funct.*

*Mater.* **28**, 1803072, doi:10.1002/adfm.201803072 (2018).

12  Erickson, D., Sinton, D. & Psaltis, D. Optofluidics for energy applications. *Nat. Photonics* **5**, 583-590, doi:10.1038/nphoton.2011.209 (2011).

13  Wunenburger, R. *et al.* Fluid flows driven by light scattering. *J. Fluid Mech.* **666**, 273-307, doi:10.1017/s0022112010004180 (2010).

14  Ashkin, A. Acceleration and Trapping of Particles by Radiation Pressure. *Phys. Rev. Lett.* **24**, 156-159, doi:10.1103/PhysRevLett.24.156 (1970).

15  Ai, X. *et al.* Photoacoustic laser streaming with non-plasmonic metal ion implantation in transparent substrates. *Opt. Express* **29**, doi:10.1364/oe.430025 (2021).

16  Yue, S. *et al.* Gold-implanted plasmonic quartz plate as a launch pad for laser-driven photoacoustic microfluidic pumps. *Proc. Natl. Acad. Sci. U. S. A.* **116**, 6580-6585, doi:10.1073/pnas.1818911116 (2019).

17  Wang, W. *et al.* Multifunctional ferrofluid-infused surfaces with reconfigurable multiscale topography. *Nature* **559**, 77-82, doi:10.1038/s41586-018-0250-8 (2018).

18  Li, A. *et al.* Programmable droplet manipulation by a magnetic-actuated robot. *Sci. Adv.* **6**, eaay5808, doi:10.1126/sciadv.aay5808 (2020).

19  Dunne, P. *et al.* Liquid flow and control without solid walls. *Nature* **581**, 58-62, doi:10.1038/s41586-020-2254-4 (2020).

20  Taylor, R. *et al.* Small particles, big impacts: A review of the diverse applications of nanofluids. *Appl. Phys.* **113**, 011301, doi:10.1063/1.4754271 (2013).

21  Fan, X., Dong, X., Karacakol, A. C., Xie, H. & Sitti, M. Reconfigurable multifunctional ferrofluid droplet robots. *Proc. Natl. Acad. Sci. U. S. A.* **117**, 27916-27926, doi:10.1073/pnas.2016388117 (2020).

22  Li, M. *et al.* Flexible magnetic composites for light-controlled actuation and interfaces. *Proc. Natl. Acad. Sci. U. S. A.* **115**, 8119-8124, doi:10.1073/pnas.1805832115 (2018).

23  Barnes, G. Rotary Curie - point heat engine. *Phys. Teach.* **24**, 204-210, doi:10.1119/1.2341985 (1986).

24  Rosensweig, R., Kaiser, R. & Miskolczy, G. Viscosity of magnetic fluid in a magnetic field. *J. Colloid Interface Sci.* **29**, 680-686, doi:10.1016/0021-9797(69)90220-3 (1969).

25  Scriven, L. & Sternling, C. The marangoni effects. *Nature* **187**, 186-188, doi:10.1038/187186a0 (1960).

26  Pshenichnikov, A. Equilibrium magnetization of concentrated ferrocolloids. *J. Magn. Magn. Mater.* **145**, 319-326, doi:10.1016/0304-8853(94)01632-1 (1995).

27  Ivanov, A. O. & Elfimova, E. A. Low-temperature magnetic susceptibility of concentrated ferrofluids: The influence of polydispersity. *J. Magn. Magn. Mater.* **374**, 327-332, doi:10.1016/j.jmmm.2014.08.067 (2015).

28  Shahsavar, A., Saghafian, M., Salimpour, M. R. & Shafii, M. B. Effect of temperature and concentration on thermal conductivity and viscosity of ferrofluid loaded with carbon nanotubes. *Heat Mass Transf.* **52**, 2293-2301, doi:10.1007/s00231-015-1743-8 (2015).

29  Liu, L. *et al.* Spinning a Liquid Wheel and Driving Surface Thermomagnetic Convection with Light. *Advanced Materials* **36**, 2306756, doi:https://doi.org/10.1002/adma.202306756 (2024).

30  Wang, L., Wang, Y., Yan, X., Wang, X. & Feng, B. Investigation on viscosity of $Fe_3O_4$ nanofluid under magnetic field. *Int. Commun. Heat Mass Transf.* **72**, 23-28, doi:10.1016/j.icheatmasstransfer.2016.01.013 (2016).

31  Ghasemi, E., Mirhabibi, A. & Edrissi, M. Synthesis and rheological properties of an iron oxide ferrofluid. *J. Magn. Magn. Mater.* **320**, 2635-2639, doi:10.1016/j.jmmm.2008.05.036 (2008).

32  Singh Mehta, J., Kumar, R., Kumar, H. & Garg, H. Convective heat transfer enhancement using ferrofluid: a review. *J. Therm. Sci. Eng. Appl.* **10**, doi:10.1115/1.4037200 (2018).

33  Love, L. J., Jansen, J. F., McKnight, T. E., Roh, Y. & Phelps, T. J. A magnetocaloric pump for microfluidic applications. *IEEE Trans. Nanobioscience* **3**, 101-110, doi:10.1109/tnb.2004.828265 (2004).

**Table of contents**

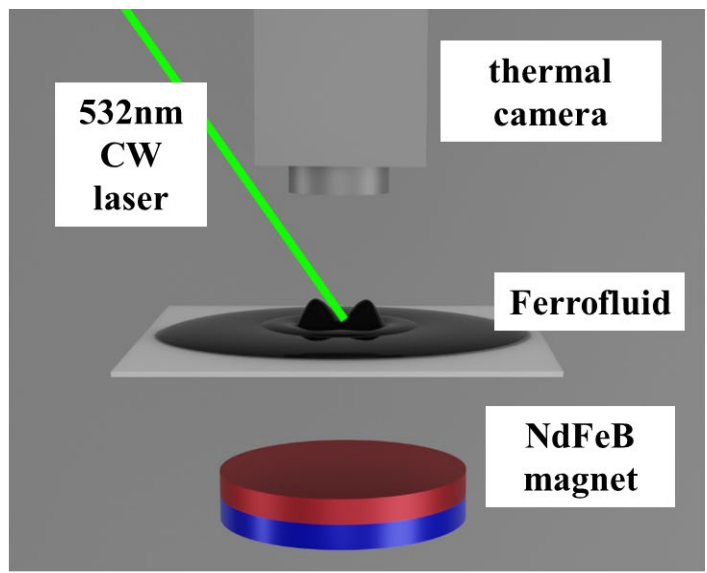

Schematic diagram of the experimental setup

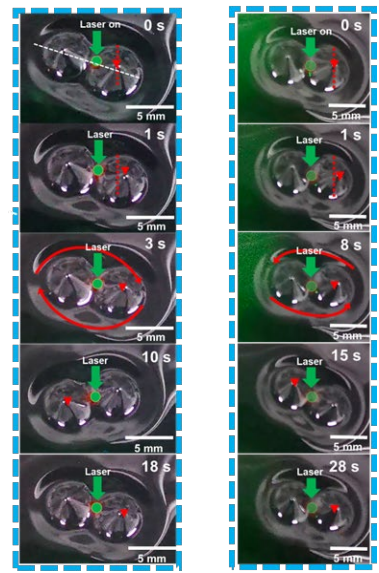

Laser spinning of ferrofluid droplets

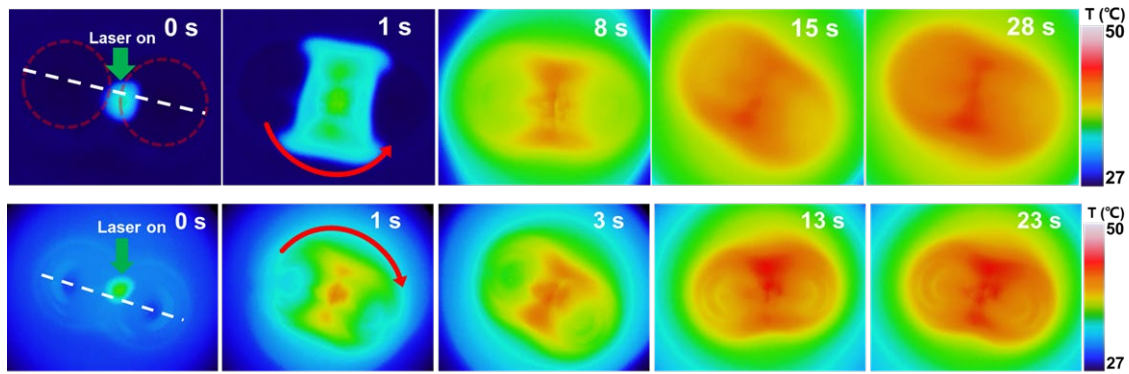

**Thermal Images of Rotating Ferrofluid**

The spin of ferrofluid droplets on the horizontal plane driven by laser is realized by destroying the force balance of the fluid and inducing asymmetric force through the photothermal effect.

Ferrofluid is harnessed using laser irradiation to generate Marangoni and thermomagnetic forces, enabling rotation due to its unique combination of liquid properties and magnetic responsiveness.